\documentclass[onecolumn,showpacs,preprintnumbers,amsmath,amssymb]{revtex4}
\usepackage[latin9]{inputenc}
\usepackage{amsmath}
\usepackage{amssymb}
\usepackage{graphicx}
\usepackage{graphics}
\usepackage{color}
\usepackage{slashed}
\usepackage{bm}
\usepackage{epstopdf}
\usepackage[unicode=true,
 bookmarks=true,bookmarksnumbered=false,bookmarksopen=false,
 breaklinks=false,pdfborder={0 0 1},backref=false,colorlinks=false]
 {hyperref}

\makeatletter

\@ifundefined{textcolor}{}
{%
 \definecolor{BLACK}{gray}{0}
 \definecolor{WHITE}{gray}{1}
 \definecolor{RED}{rgb}{1,0,0}
 \definecolor{GREEN}{rgb}{0,1,0}
 \definecolor{BLUE}{rgb}{0,0,1}
 \definecolor{CYAN}{cmyk}{1,0,0,0}
 \definecolor{MAGENTA}{cmyk}{0,1,0,0}
 \definecolor{YELLOW}{cmyk}{0,0,1,0}

}

\makeatother

\begin{document}

\title{Dirichlet boundary condition for the Lee-Wick-like scalar model}

\author{L.H.C. Borges$^{1}$}
\email{luizhenriqueunifei@yahoo.com.br}

\author{A.A. Nogueira$^{2}$}
\email{andsogueira@hotmail.com}

\author{E.H. Rodrigues$^{3}$}
\email{eversonhr@gmail.com}

\author{F.A. Barone$^{3}$}
\email{fbarone@unifei.edu.br}

\affiliation{$^{1}$Universidade Federal do ABC, Centro de Ci\^encias Naturais e Humanas,
Rua Santa Ad\'elia, 166, 09210-170, Santo Andr\'e, SP, Brazil}

\affiliation{$^{2}$Fazendinha, Zona Rural, 37460-000, Passa Quatro, MG, Brazil}

\affiliation{$^{3}$IFQ - Universidade Federal de Itajub\'a, Av. BPS 1303, Pinheirinho, Caixa Postal 50, 37500-903, Itajub\'a, MG, Brazil}

\begin{abstract}
Lee-Wick-like scalar model near a Dirichlet plate is considered in this work. The modified propagator for the scalar field due to the presence of a Dirichlet boundary is computed, and the interaction between the plate and a point-like scalar charge is analysed. The non-validity of the image method is investigated and the results are compared with the corresponding ones obtained for the Lee-Wick gauge field and for the standard Klein-Gordon field.
\end{abstract}

\maketitle

\section{\label{I}Introduction}

One of the most remarkable features of models with higher order derivatives  concerns on the fact that they tame some ultraviolet divergences in field theories, both at classical as well as at quantum levels. This is the main reason why these models are intensively studied in the literature. The simplest model of this type was proposed by B. Podolsky and T. Lee and G. Wick \cite{LW1,LW2,LW3,LW4,LW5,LW6}, where the self energy for a point-like charge in $3+1$ dimensions is finite \cite{LW7,LW8,LW9,LW11}. Nowdays, models similar to the one proposed  B. Podolsky and T. Lee and G. Wick are called Lee-Wick models.

Some aspects of Lee-Wick models have been investigated, for example, in interactions between external sources \cite{FABAAN2}, issues related with the ghosts problems \cite{Tonsk,Complex,Fakeon}, radiative corrections \cite{radiative}, propagation of waves \cite{waves}, dual symmetry \cite{Brandt}, phenomenological and cosmological implications \cite{ph1,ph2,ph3,ph4,ph5,ph6,c1,c2,c3}, the connection with Pauli-Villars regularization \cite{Thibes}, and so on. In fact, there is a vast literature concerning Lee-Wick models.

It is well known that field theories with the presence of boundary conditions is a subject with a wide range of application in several branches of physics. We highlight, the phenomena related to the presence of material boundaries in field theories \cite{Casimir,Mostepanenko,BordagLivro,Milton,MiltonLivro,CasimirKR,Ravndal,FTAB,Scharnhorst,ScharnhorstBarton,ScharnhorstNos,VacuoQED,Emissaoexp,MagneticoVacuoEscalar,MagneticoVacuoQED,AtomoParede,AtomoCunha,FabricioTese}, just to mention a few examples. An alternative approach to this kind of problem is the study of models with fields coupled to external potencials defined along surfaces \cite{Bord,Jaff,Milton,GTFABFEB,FABFEB1,FABFEB2,LHCBAFFFAB,LHCBFABplate,FABAANBMP}.

Regarding the presence of boundary conditions in the Lee-Wick Electrodynamics, we can mention the study of the Casimir effect \cite{FabAnd,Blaz} and the Lee-Wick Electrodynamics in the vicinity of a perfectly conducting plate, which is an example of a linear field theory where the image method is not valid \cite{Amaster,FABAAN1}. However, the Lee-Wick-like scalar model have not been considered in the presence of boundary conditions up to now, as far as the authors know. A natural question which arises in this scenario concerns on the modifications in which the Lee-Wick scalar propagator undergoes due to the presence of a single Dirichlet plane, and the influence of this kind of surface on the dynamics of point-like field sources. A related question to this subject regards on the fact that some divergences in theories with higher order derivatives, but without boundary conditions, are naturally controlled. Are these divergences still controlled with the presence of boundary conditions? Investigations of this type have not been explored in the literature as it should.

This paper is devoted to this subject, where we investigate some aspects of the Lee-Wick-like scalar model in the vicinity of a single Dirichlet plane. More specifically, in Sect. \ref{II} we compute the propagator for the Lee-Wick scalar field in the presence of a Dirichlet plane. In Sect. \ref{III} from the propagator founded previously, we calculate the interaction energy as well as the interaction force between the point-like scalar charge and the Dirichlet plane. We obtain exact results only for some special situations and we perform numerical analysis for more general cases. We compare the interaction forces with the ones obtained in the free theory (theory without the Dirichlet plane) and verify that the image method is not valid in our model for the Dirichlet boundary condition. We also compare the results obtained along the paper with the corresponding ones computed for the abelian Lee-Wick model (for the gauge field) as well as with the ones obtained for the standard Klein-Gordon field theory. We show that the interaction energy between the Dirichlet plane and a point-like charge is finite, even in the limit when the charge is placed near on the plane. In Sect. \ref{IV} we consider the case where the charge is placed on the Dirichlet mirror. Finally, Sect. \ref{V} is devoted to our final remarks and conclusions.

Throughout the paper we shall deal with a scalar model with higher order derivatives, in a 3+1 dimensional space-time. It is used Minkowski coordinates with the diagonal metric with signature (+, -, -, -).

\section{\label{II} The modified propagator}

The Lagrangian density for the real Lee-Wick-like scalar model is given by  
\begin{eqnarray}
\label{Model}
{\cal{L}}=\frac{1}{2}\partial_{\mu}\phi\partial^{\mu}\phi +\frac{1}{2}\partial_{\mu}\phi\frac{\Box}{m^{2}}\partial^{\mu}\phi-\frac{1}{2}M^{2}\phi^{2}+J\phi \ ,
\end{eqnarray}
where $\phi$ is a real scalar field, $M$ stands for mass of the scalar field, $J$ is the external source and $m$ is a parameter with mass dimension.

From now on, she shall name $m$ and $M$ as Podolsy mass and Klein-Gordon mass, respectively. 

The propagator for the model (\ref{Model}) reads
\begin{eqnarray}
\label{Prop}
D\left(x,y\right)=\int\frac{d^{4}p}{(2\pi)^{4}}\frac{m^{2} \ e^{-ip\cdot\left(x-y\right)}}{p^{4}-m^{2}p^{2}+M^{2}m^{2}} \ ,
\end{eqnarray}
where the free propagator $D\left(x,y\right)$ is the inverse of the kinetic operator ${\cal{O}}$,
\begin{eqnarray}
\label{Operator1}
{\cal{O}}=\Box +\frac{\Box^{2}}{m^{2}}+M^{2} \ ,
\end{eqnarray}
in the sense that,
\begin{eqnarray}
\label{Operator2}
{\cal{O}}D\left(x,y\right)=\delta^{4}\left(x-y\right) \ .
\end{eqnarray}

From Eq. (\ref{Prop}) one can show that the model (\ref{Model}) exhibits two massive poles for momentum square, namely
\begin{eqnarray}
\label{Modes}
m_{\pm}^{2}=\frac{m^{2}}{2}\left(1\pm\sqrt{1-\frac{4M^{2}}{m^{2}}}\right) \ ,
\end{eqnarray}

In order to avoid ghost modes, one must impose the following condition
\begin{eqnarray}
\label{Condition}
0\leq\frac{4M^{2}}{m^{2}}\leq 1 \ .
\end{eqnarray}

We can notice that if $0 < (4M^{2}/m^{2}) < 1$ we have two field modes with different non vanishing massive poles, $m_{+}$ and $m_{-}$. In this situation the propagator can be rewritten in the following way
\begin{eqnarray}
\label{prop2}
D\left(x,y\right)=\frac{1}{\sqrt{1-\frac{4M^{2}}{m^{2}}}}\int\frac{d^{4}p}{(2\pi)^{4}}\left(\frac{1}{p^{2}-m_{+}^{2}}-\frac{1}{p^{2}-m_{-}^{2}}\right)e^{-ip\cdot\left(x-y\right)}
\end{eqnarray}

The propagator (\ref{prop2}) exhibits a singular configuration, in which we have two terms with opposite signs, each one with a different massive pole. This kind of propagator leads to a wide range of physical phenomena and is richer than the corresponding one found for the Lee-Wick gauge field, where we have just one massive pole and one massless pole. The relevance of this kind of propagator was explored in many contexts, such as of the Higgs and Debye screening \cite{BGNP}, in order to understand how the concept of mass could be useful to handle divergences in field theories not only in infra-red, but also in ultraviolet regime. It was also considered in the regularization of the self energy of point-like charges \cite{LW7}, where we can have a finite self energy in a wider range of situations in comparison with the Lee-Wick Electrodynamics.  

In this paper we shall investigate the role played by these two massive poles when the field is submitted to boundary conditions. Previously in the literature, it was considered the Lee-Wick gauge field submitted to boundary conditions imposed by a perfectly conducting plate. As far as the authors know, results concerning a field theory with two massive poles and in the presence of boundary conditions were not considered in the literature up to now.

From now on, we shall consider the field described by the model (\ref{Model}) in the presence of a single Dirichlet plane. Namely, we shall impose that the field is equal to zero along a given plane. Without loss of generality, we take a coordinate system where the Dirichlet plane is perpendicular to the $x^{3}$ axis and placed at $x^{3}=a$. So, the Dirichlet boundary condition imposed on the field reads 
\begin{eqnarray}
\label{Dirich}
\phi\left(x\right)\mid_{x^{3}=a}=0 \ ,
\end{eqnarray}
where the sub-index means that the boundary condition is taken on the plane $x^{3}=a$.

We must obtain the functional generator for the scalar field submitted to the boundary condition (\ref{Dirich}). For this task we use the functional formalism proposed in \cite{BRW} and employed in \cite{FABAANBMP,Helder,BBH}. We start by writing the functional generator as follows
\begin{eqnarray}
\label{fgen1}
Z_{C}\left[J\right]=\int {\cal{D}}\phi_{C} \ e^{i\int d^{4}x \ \cal{L}} \ ,
\end{eqnarray}
where the sub-index $C$ means that we are integrating out in all the field configurations which satisfy the condition (\ref{Dirich}). This restriction is attained by introducing  a functional delta that is not equal to zero only where the condition (\ref{Dirich}) is satisfied, as follows
\begin{eqnarray}
\label{fgen2}
Z_{C}\left[J\right]=\int {\cal{D}}\phi \ \delta\left[\phi\left(x\right)\mid_{x^{3}=a}\right] \ e^{i\int d^{4}x \ \cal{L}} \ .
\end{eqnarray}

The delta functional present in (\ref{fgen2}) has the Fourier representation
\begin{eqnarray}
\label{fgen3}
\delta\left[\phi\left(x\right)\mid_{x^{3}=a}\right] =\int {\cal{D}}B\exp\left[-i
\int d^{4}x\ \delta\left(x^{3}-a\right)B\left(x_{\parallel}\right)\phi
\left(x\right)\right] \ ,
\end{eqnarray}
where $B\left(x_{\parallel}\right)$ is an auxiliary scalar field and $x_{\parallel}^{\mu}=\left(x^{0},x^{1},x^{2}\right)$ means that we have only the coordinates parallel to the plane.

Substituting expression (\ref{fgen3}) in (\ref{fgen2}), we have that
\begin{eqnarray}
\label{fgen4}
Z_{C}\left[J\right]=\int {\cal{D}}\phi {\cal{D}}B\ e^{i\int d^{4}x \ \cal{L}}\exp\left[-i
\int d^{4}x\ \delta\left(x^{3}-a\right)B\left(x_{\parallel}\right)\phi
\left(x\right)\right] \ .
\end{eqnarray}

In order to write the above integral in a convenient form, we perform the following translation in the $\phi$  field 
\begin{eqnarray}
\label{Translation}
\phi\left(x\right)\rightarrow\phi\left(x\right)-\int d^{4}y \ \delta\left(y^{3}-a\right)D\left(x,y\right)B\left(y_{\parallel}\right) \ ,
\end{eqnarray}
which has unitary jacobian and enables us to write (\ref{fgen4}) as follows
\begin{eqnarray}
\label{fgen5}
Z_{C}\left[J\right]=Z\left[J\right]{\bar{Z}}\left[J\right] \ ,
\end{eqnarray}
where $Z\left[J\right]$ is the usual functional generator for the scalar field
\begin{eqnarray}
\label{fgen6}
Z\left[J\right]=\int{\cal{D}}\phi\ e^{i\int d^{4}x \ \cal{L}}
=Z\left[0\right]\exp\left[\frac{i}{2}\int d^{4}x \ d^{4}y \ J\left(x\right)D\left(x,y\right)J\left(y\right)\right] \ ,
\end{eqnarray}
and ${\bar{Z}}\left[J\right]$ is a contribution due to the auxiliary scalar field $B$  
\begin{eqnarray}
\label{fgen7}
{\bar{Z}}\left[J\right]=\int{\cal{D}}B\exp\left[-i\int d^{4}y \ \delta
\left(y^{3}-a\right)I\left(y\right)B\left(y_{\parallel}\right)\right] \nonumber\\
\times\exp\left[\frac{i}{2}\int d^{4}x \ d^{4}y \ \delta\left(x^{3}-a\right)
\delta\left(y^{3}-a\right)B\left(x_{\parallel}\right)D\left(x,y\right)
B\left(y_{\parallel}\right)\right] \ ,
\end{eqnarray}
where we identified
\begin{eqnarray}
\label{defi1}
I\left(y\right)=\int d^{4}x \ D\left(x,y\right)J\left(x\right)  \ . 
\end{eqnarray}

Substituting (\ref{defi1}) and (\ref{Prop}) into (\ref{fgen7}), defining  the momentum parallel to the plane $p_{\parallel}^{\mu}=\left(p^{0},p^{1},p^{2}\right)$, the quantities $\Gamma=\sqrt{p_{\parallel}^{2}-m_{+}^{2}} \ ,  \ L=\sqrt{p_{\parallel}^{2}-m_{-}^{2}}$, and the parallel metric
\begin{eqnarray}
\label{etap}
\eta_{\parallel}^{\mu\nu}=\eta^{\mu\nu}-\eta_{\ 3}^{\mu}\eta^{\nu 3} \ ,
\end{eqnarray}
and using the fact that \cite{FABAAN1,GTFABFEB,LHCBAFFFAB} 
\begin{eqnarray}
\label{int}
\int \frac{dp^{3}}{2\pi}\frac{e^{i p^{3}\left(x^{3}-y^{3}\right)}}
{p^{2}-m_{+}^{2}}=-\frac{i}{2\Gamma} \ e^{i\Gamma\mid x^{3}-y^{3}\mid} \ , \   \int \frac{dp^{3}}{2\pi}\frac{e^{i p^{3}\left(x^{3}-y^{3}\right)}}
{p^{2}-m_{-}^{2}}=-\frac{i}{2L} \ e^{iL\mid x^{3}-y^{3}\mid} \ ,
\end{eqnarray}
where $p^{3}$ stands for the momentum perpendicular to the plane, we can write Eq. (\ref{fgen7}) in the form
\begin{eqnarray}
\label{fgen8}
{\bar{Z}}\left[J\right]={\bar{Z}}\left[0\right]\exp\left[\frac{i}{2}\int d^{4}x 
\ d^{4}y \ J\left(x\right){\bar{D}}\left(x,y\right)J\left(y\right)\right] \ ,
\end{eqnarray}
where we defined the function
\begin{eqnarray}
\label{Propplane}
{\bar{D}}\left(x,y\right)&=&-\frac{i}{2}\frac{1}{\sqrt{1-\frac{4M^{2}}{m^{2}}}}\int \frac{d^{3}p_{\parallel}}
{\left(2\pi\right)^{3}}\frac{1}{\frac{1}{L}-\frac{1}{\Gamma}} \exp\left[{-i p_{\parallel}\cdot\left(x_{\parallel}
-y_{\parallel}\right)}\right]\nonumber\\
&
&\times
\left(\frac{e^{iL\mid x^{3}-a\mid}}{L}-\frac{e^{i\Gamma\mid x^{3}-a\mid}}{\Gamma}\right)
\left(\frac{e^{iL\mid y^{3}-a\mid}}{L}-\frac{e^{i\Gamma\mid y^{3}-a\mid}}{\Gamma}\right) \ .
\end{eqnarray}

With the aid of (\ref{fgen6}), and using expression (\ref{fgen8}), Eq. (\ref{fgen5}) becomes
\begin{eqnarray}
\label{fgen9}
Z_{C}\left[J\right]=Z_{C}\left[0\right]\exp\left[\frac{i}{2}
\int d^{4}x \ d^{4}y \ J\left(x\right)\left(D
\left(x,y\right)+{\bar{D}}\left(x,y\right)\right)J
\left(y\right)\right] \ .
\end{eqnarray}

From Eq. (\ref{fgen9}), we can identify the propagator of the theory in the presence of a Dirichlet plane as follows
\begin{eqnarray}
\label{Proptotal}
D_{C}\left(x,y\right)=D\left(x,y\right)+{\bar{D}}\left(x,y\right) \ .
\end{eqnarray}

The propagator (\ref{Proptotal}) is composed by the sum of the free propagator (\ref{prop2}) with the correction (\ref{Propplane}) which accounts for the presence of the Dirichlet plane. In the limit $m\rightarrow\infty$ the propagator (\ref{Propplane}) reduces to the same one as that found with the standard Klein-Gordon field theory in the presence of a Dirichlet plane.

We can check the results by taking the particular classical solution for the scalar field generated by an external source,
\begin{eqnarray}
\label{Field}
\phi\left(x\right)=\int d^{4}y \  D_{C}\left(x,y\right)J\left(y\right) \ .
\end{eqnarray}

Once $D_{C}\left(x,y\right)\mid_{x^{3}=a}=0 $, we can show that the solution (\ref{Field}) really satisfies the boundary condition (\ref{Dirich}).

\section{\label{III} Particle--plane interaction}

In this section we consider the interaction energy between a point-like external source and the Dirichlet plane, which is given by \cite{GTFABFEB,LHCBAFFFAB}
\begin{eqnarray}
\label{Energy}
E=-\frac{1}{2T}\int d^{4}x \  d^{4}y \  J\left(x\right){\bar{D}}\left(x,y\right)J\left(y\right) \ .
\end{eqnarray}

With no loss of generality, and for simplicity, we choose
a point-like scalar charge located at position  ${\bf{b}} = \left(0, 0, b\right)$. The corresponding external source reads
\begin{eqnarray}
\label{Source}
J\left(x\right)=q\delta^{3}\left({\bf{x}}
-{\bf{b}}\right) \ ,
\end{eqnarray}
where the parameter $q$ can be seen as a kind of scalar charge.

Substituting expressions (\ref{Source}) and (\ref{Propplane}) in (\ref{Energy}), carrying out the the integrals in $d^{3}{\bf{x}},d^{3}{\bf{y}},dx^{0},dp^{0},dy^{0}$, and then, performing some simple manipulations, we arrive at
\begin{eqnarray}
\label{Energy2}
E_{PC}&=&\frac{q^{2}}{4}\frac{1}{\sqrt{1-\frac{4M^{2}}{m^{2}}}}\int \frac{d^{2}{\bf{p}}_{\parallel}}
{\left(2\pi\right)^{2}}\frac{\sqrt{{\bf{p}}_{\parallel}^{2}+m_{+}^{2}}\sqrt{{\bf{p}}_{\parallel}^{2}+m_{-}^{2}}}{\sqrt{{\bf{p}}_{\parallel}^{2}+m_{+}^{2}}-\sqrt{{\bf{p}}_{\parallel}^{2}+m_{-}^{2}}}\nonumber\\
&
&\times
\left(\frac{e^{-\sqrt{{\bf{p}}_{\parallel}^{2}+m_{-}^{2}}R}}{\sqrt{{\bf{p}}_{\parallel}^{2}+m_{-}^{2}}}-\frac{e^{-\sqrt{{\bf{p}}_{\parallel}^{2}+m_{+}^{2}}R}}{\sqrt{{\bf{p}}_{\parallel}^{2}+m_{+}^{2}}}\right)^{2} \ ,
\end{eqnarray}
where $R=\mid a-b\mid $  stands for the distance between the plane and the charge. The sub-index $PC$ means that we have the interaction energy between the Dirichlet plane and the charge.

Eq. (\ref{Energy2}) can be simplified by using polar coordinates and integrating out in the angular variables, 
\begin{eqnarray}
\label{Energy3}
E_{PC}&=&\frac{q^{2}}{8\pi}\frac{m_{+}^{2}+m_{-}^{2}}{(m_{+}^{2}-m_{-}^{2})^2}
\int_{0}^{\infty}dp \  p\sqrt{p^{2}+m_{+}^{2}}\sqrt{p^{2}+m_{-}^{2}}\left(\sqrt{p^{2}+m_{+}^{2}}+\sqrt{p^{2}+m_{-}^{2}}\right)\nonumber\\
&
&\times\left(\frac{e^{-2\sqrt{p^{2}+m_{-}^{2}}R}}{p^{2}+m_{-}^{2}}-2\frac{e^{-\left(\sqrt{p^{2}+m_{-}^{2}}+\sqrt{p^{2}+m_{+}^{2}}\right)R}}{\sqrt{p^{2}+m_{-}^{2}}\sqrt{p^{2}+m_{+}^{2}}}+\frac{e^{-2\sqrt{p^{2}+m_{+}^{2}}R}}{p^{2}+m_{+}^{2}}\right) \ .
\end{eqnarray}

The result (\ref{Energy3}) is exact and gives the interaction energy between a Dirichlet plane and a point-like scalar charge for the model (\ref{Model}). It is important to point out that in the limit $m\rightarrow\infty$ the Eq. (\ref{Energy3}) reproduces the same result founded in Ref. \cite{GTFABFEB}.

We could not solve the integral on the right hand side of (\ref{Energy3})  analytically in the general case. In order to have a better insight on the meaning of above expression, at first, we analyse  two special cases where it is possible to solve the integral in (\ref{Energy3}).

In the first case we take $M=0$ in Eq. (\ref{Energy3}). In this situation, from Eq. (\ref{Modes}), we have 
$m_ {-}=0$ and $m_{+}=m$. So, after some manipulations the interaction energy (\ref{Energy3}) becomes 
\begin{eqnarray}
\label{Energy4}
E_{PC}\left(M=0\right)&=&\frac{q^{2}m}{8\pi}\int_{0}^{\infty}dp \  p^{2}\left[\left(p^{2}+1\right)+p\sqrt{p^{2}+1}\right]\nonumber\\
&
&\times\left(\frac{e^{-2pmR}}{p^{2}}-2\frac{e^{-\left(p+\sqrt{p^{2}+1}\right)mR}}{p\sqrt{p^{2}+1}}+\frac{e^{-2\sqrt{p^{2}+1}mR}}{p^{2}+1}\right) \ .
\end{eqnarray}

Using the fact that \cite{FABAAN1}
\begin{eqnarray}
\label{Integrals}
\int_{0}^{\infty}dp \left[\left(p^{2}+1\right)+p\sqrt{p^{2}+1}\right]e^{-2pmR}&=&\frac{1+2\left(mR\right)^{2}}{4\left(mR\right)^{3}}+ \frac{\pi}{4mR}\left[Y_{0}\left(2mR\right)-SH_{0}\left(2mR\right)\right] \nonumber\\
&
&+\frac{\pi}{4\left(mR\right)^{2}}\left[SH_{1}\left(2mR\right)-Y_{1}\left(2mR\right)\right] \ ,\nonumber\\
\int_{0}^{\infty}dp \ p^{2}\left(1+\frac{p}{\sqrt{p^{2}+1}}\right)e^{-2\sqrt{p^{2}+1}mR}&=&\frac{K_{0}\left(2mR\right)}{2mR}+\frac{K_{1}\left(2mR\right)}{2\left(mR\right)^{2}}+\frac{e^{-2mR}}{4\left(mR\right)^{3}}\left(1+2mR\right) \ , \nonumber\\
-2\int_{0}^{\infty}dp \ p\left(\sqrt{p^{2}+1}+p\right)e^{-\left(p+\sqrt{p^{2}+1}\right)mR}&=& e^{-mR}\left(\frac{1}{2}-\frac{1}{2mR}-\frac{1}{\left(mR\right)^{2}}-\frac{1}{\left(mR\right)^{3}}\right)\nonumber\\
&
&-\frac{1}{2}\left(mR\right)Ei\left(1,mR\right) \ ,
\end{eqnarray}
where $Y, SH, K$ stand for the Bessel function of second kind, the Struve function and the K-Bessel function 
respectively, and $Ei$ is the exponential integral function, which is defined by \cite{Arfken}
\begin{equation}
\label{Ei1}
Ei(n,s)=\int_{1}^{\infty}\frac{e^{-ts}}{t^{n}}\ dt\ \ \,\ \ \ \Re(s)>0\ ,\ n=0,1,2,\cdots\ ,
\end{equation}
the energy reads
\begin{eqnarray}
\label{Energy5}
E_{PC}\left(M=0\right)&=&\frac{q^{2}}{16\pi R}\left[1+\Delta\left(mR\right)\right] \ ,
\end{eqnarray}
where
\begin{eqnarray}
\label{deltamR}
\Delta\left(mR\right)&=&mR\Biggl[\frac{K_{0}\left(2mR\right)}{mR}+\frac{K_{1}\left(2mR\right)}{\left(mR\right)^{2}}+\frac{e^{-2mR}}{2\left(mR\right)^{3}}+\frac{e^{-2mR}}{\left(mR\right)^{2}}-\left(mR\right)Ei\left(1,mR\right)\nonumber\\
&
&+\frac{1}{2\left(mR\right)^{3}}+\frac{\pi}{2mR}\left[Y_{0}\left(2mR\right)-SH_{0}\left(2mR\right)\right]+\frac{\pi}{2\left(mR\right)^{2}}\left[SH_{1}\left(2mR\right)-Y_{1}\left(2mR\right)\right]\nonumber\\
&
&+e^{-mR}\left(1-\frac{1}{mR}-\frac{2}{\left(mR\right)^{2}}-\frac{2}{\left(mR\right)^{3}}\right)
\Biggr] \ .
\end{eqnarray}

In Eq. (\ref{Energy5}) the first term on the right hand side is the plane-charge interaction obtained in standard Klein-Gordon field theory. The second term is a correction imposed by the parameter $m$ which falls when $mR$ increases.

It is important to mention that the integrals (\ref{Integrals}) are valid just for $mR\not=0$. When $R=0$ all the three integrals are divergent and the case where $R$ must be treated carefully. Is corresponds to take the charge plced on the Dirichlet plate and we shall leave this situation to be analized lastly. 

The interaction force between the Dirichlet plane and the scalar charge is given by
\begin{eqnarray}
\label{Force1}
F_{PC}\left(M=0\right)=-\frac{\partial E_{PC}\left(M=0\right)}{\partial R}=\frac{q^{2}}{16\pi R^{2}}\left[1+\Delta\left(mR\right)-mR \ \Delta'\left(mR\right)\right] \ ,
\end{eqnarray}
where the prime denotes derivative of $\Delta$ with respect to its argument. 

In Eq. (\ref{Force1}) the first term between brackets is the interaction force between the point charge and the Dirichlet plane for the massless scalar field obtained in the free Klein-Gordon field theory. It is the usual Coulombian force with an overall minus sign, between the scalar charge and its image placed at a distance $2R$ apart each other. The remaining term in (\ref{Force1}) is an m-dependent correction which falls when $mR$ increases. So, we notice that the quantity inside brackets is always positive (since $m$ is a large quantity, the correction term is much smaller than the Coulombian one), what means that the force is always repulsive.

It is interesting to notice that in the limit $R\rightarrow 0$ both energy and force are finite, as follows
\begin{eqnarray}
\label{R0M0}
\lim_{R\rightarrow 0}E_{PC}\left(M=0\right)=\frac{mq^{2}}{8\pi}\  , \  \   \lim_{R\rightarrow 0}F_{PC}\left(M=0\right)=\frac{3m^{2}q^{2}}{32\pi} \ .
\end{eqnarray}
This fact is due to the Lee-Wick term in the model (\ref{Model}) and shows that the models with higher order derivatives improve renormalization properties and tame ultraviolet divergences present in the
scalar field theory with Dirichlet boundary conditon, since in the standard Klein-Gordon field theory $(m\rightarrow\infty)$ the corresponding results to (\ref{R0M0}) are divergent in the limit $R\rightarrow 0$.

It is worth mentioning that the interaction force (\ref{Force1}) is equivalent to that one obtained in the Reference \cite{FABAAN1} (with an overall minus sign), for the gauge field, where we have the interaction between a point-like charge and a perfectly conducting plate mediated by electromagnetic Lee-Wick  model. This result shows that the connection between the scalar field in the presence of a Dirichlet plane and the gauge field in the presence of a perfectly conducting plate still remains when we have the corresponding Lee-Wick terms in both theories. It is really a surprisingly result because in both theories we can define a dimensionless parameter $mR$ and the functional form for the energy cannot be foreseen before the calculations are performed. 

In Eq. (43) of Ref. \cite{FABAAN2} it is calculated the interaction energy between two point-like scalar charges for the model (\ref{Model}) (theory without the Dirichlet plane). Putting $M=0 \ (m_{-}=0, \ m_{+}=m)$ in this expression, we obtain the interaction energy for the massless scalar field with higher order derivatives.  The corresponding interaction force in the special case where we have two opposite charges, $\sigma_{1}=q,\sigma_{2}=-q$, placed at a distance $a=2\mid R\mid$ apart is given by
\begin{eqnarray}
\label{Force2}
F_{CC}\left(M=0\right)=\frac{q^{2}}{16\pi R^{2}}\left(1-e^{-2mR}-2mR \ e^{-2mR}\right) \ .
\end{eqnarray}
The sub-index $CC$ means that we have the interaction between two point-like charges.

It is straightforward to verify that the force (\ref{Force2}) is different from (\ref{Force1}). So, it is interesting to notice that the image method is not valid for the point-like charge in the scalar theory (\ref{Model}) for the Dirichlet boundary condition (\ref{Dirich}). A similar situation occurs in the electromagnetic Lee-Wick model with the presence of a perfectly conducting plate \cite{FABAAN1}. 

The second case in which the integral (\ref{Energy3}) can be solved exactly it he one where $M=m/2$. It is the maximum value that the mass of the scalar field can assume without the appearance of ghosts, what can be seen from the conditions (\ref{Condition}).

Taking the limit $\left(1-\frac{4M^{2}}{m^{2}}\right)\rightarrow 0$ in Eq. (\ref{Energy3}), we have the expression 
\begin{eqnarray}
\label{Energy4m}
E_{PC}\left(M=m/2\right)&=&\frac{q^{2}m^{3}}{16\pi\sqrt{2}}\int_{1}^{\infty}dp \ e^{-\sqrt{2}pmR}\Biggl(\frac{2}{m^{2}p^{2}}+\frac{2R\sqrt{2}}{mp}\nonumber
+R^{2}\Biggr)\nonumber\\
&=&\frac{mq^{2}}{32\pi}e^{-\sqrt{2}mR}\left(2\sqrt{2}+mR\right) \ .
\end{eqnarray}

Equation (\ref{Energy4m}) gives the exact expression for the interaction energy between a point-like scalar charge and the Dirichlet plane for the special case where $M=m/2$. In the limit $R\rightarrow 0$, we have
\begin{eqnarray}
\label{enrlim0}
\lim_{R\rightarrow 0}E_{PC}\left(M=m/2\right)=\frac{mq^{2}}{8\pi\sqrt{2}} \ ,
\end{eqnarray}
which is a finite expression.

The interaction force reads
\begin{eqnarray}
\label{Force4m5}
F_{PC}\left(M=m/2\right)=-\frac{\partial }{\partial R}E_{PC}\left(M=m/2\right)=\frac{m^{2}q^{2}}{32\pi}e^{-\sqrt{2}mR}\left(3+\sqrt{2} \ mR\right) \ .
\end{eqnarray}

The above force is repulsive and in the limit $R\rightarrow 0$ is finite
\begin{eqnarray}
\label{ForceR04m}
\lim_{R\rightarrow 0}F_{PC}\left(M=m/2\right)=\frac{3m^{2}q^{2}}{32\pi} \ .
\end{eqnarray}

From Eqs. (\ref{R0M0}) and (\ref{ForceR04m}) we can notice that in the limit $R\rightarrow 0$ the force for both cases, $M=0$ and $M=m/2$, are equal to each order. 

Proceeding in the same way as in the previous case, it is simple to verify from Eq's  (\ref{Energy4m}) and from Eq. (45) of Ref. \cite{FABAAN2} that the image method is not valid in the case where $M=m/2$ for the Dirichlet boundary condition (\ref{Dirich}).

\begin{figure}[!h]
 \centering
   \includegraphics[scale=0.37]{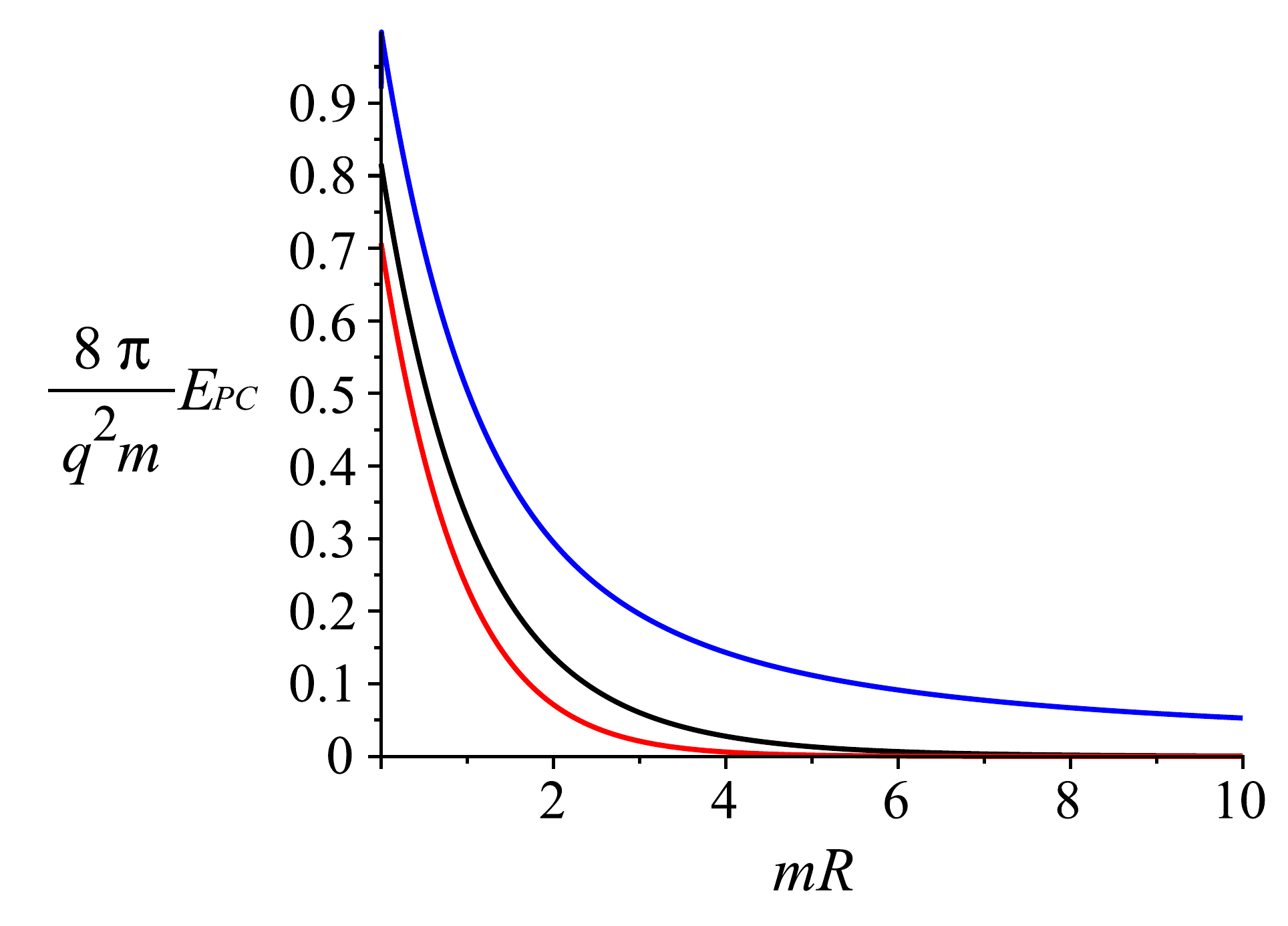} 
   \caption{Energy (\ref{Energy3}) multiplied by $8\pi/(q^{2}m)$ for: $M=0$ - blue line, $M=m/4$ - black line, and $M=m/2$ - red line.}
  \label{energias}
\end{figure}

In figure (\ref{energias}) we have a plot for the energy (\ref{Energy3}) multiplied by $8\pi/(q^{2}m)$ in three different situations, all of them with $m$ taken as fixed. In the black line, we have the maximum value for $M=m/2$, what corresponds to eq. (\ref{Energy4m}). In the red line, we have $M=m/4$ and the blue line is the case where $M=0$, the minimum value for $M$, whose energy is given by equation (\ref{Energy5}). We can see that the energy decreases monotonically when the distance $R$ increases. It increases when we decrease the values of $M$ and attains its minimal configuration for $M=m/2$. In addition, for $M$ and $m$ fixed,the energy exhibits a maximum value in the limit $R\to0$.

\begin{figure}[!h]
 \centering
   \includegraphics[scale=0.37]{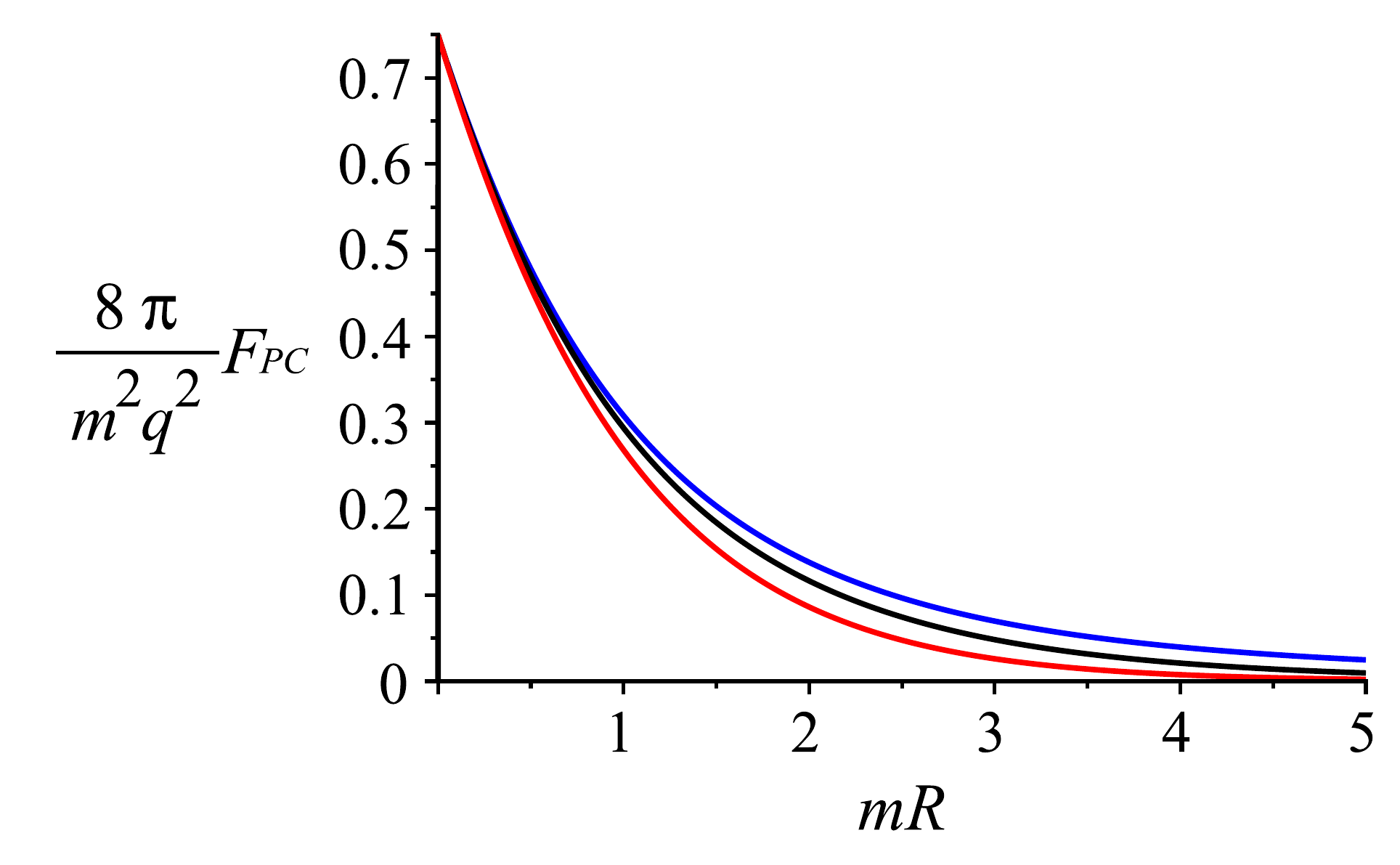}  
   \caption{Force between the Dirichlet plane and the point charge multiplied by $8\pi/(q^{2}m^{2})$ for: $M=0$ - blue line, $M=m/4$ - black line, and $M=m/2$ - red line.}
  \label{forcas}
\end{figure}

The force between the Dirichlet plane and the scalar charge is obtained by differentiating the energy (\ref{Energy3}) with respect to $R$ (with an overall minus signl). In the specific cases where $M=0$ and $M=m/2$, we were able calculate the force exactly in Eq's (\ref{Force1}) and (\ref{Force4m5}). In the figure (\ref{forcas}) we have a plot for the forces multiplied by $8\pi/(q^{2}m^{2})$ with $M=0$, $M=m/4$ and $M=m/2$ in blue, red and black, respectively. By fixing $m$, we can see that the forces are positive for all configurations, what means repulsive interactions. In addition, the three curves goes to the same value $3/4=0.75$ in the limit $R\to0$. For the maximum and minimum values, $M=m/2$ and $M=0$, we had been known this fact from (\ref{R0M0}) and (\ref{ForceR04m}). For $M=m/4$ we have just a numerical result.

\begin{figure}[!h]
 \centering
   \includegraphics[scale=0.53]{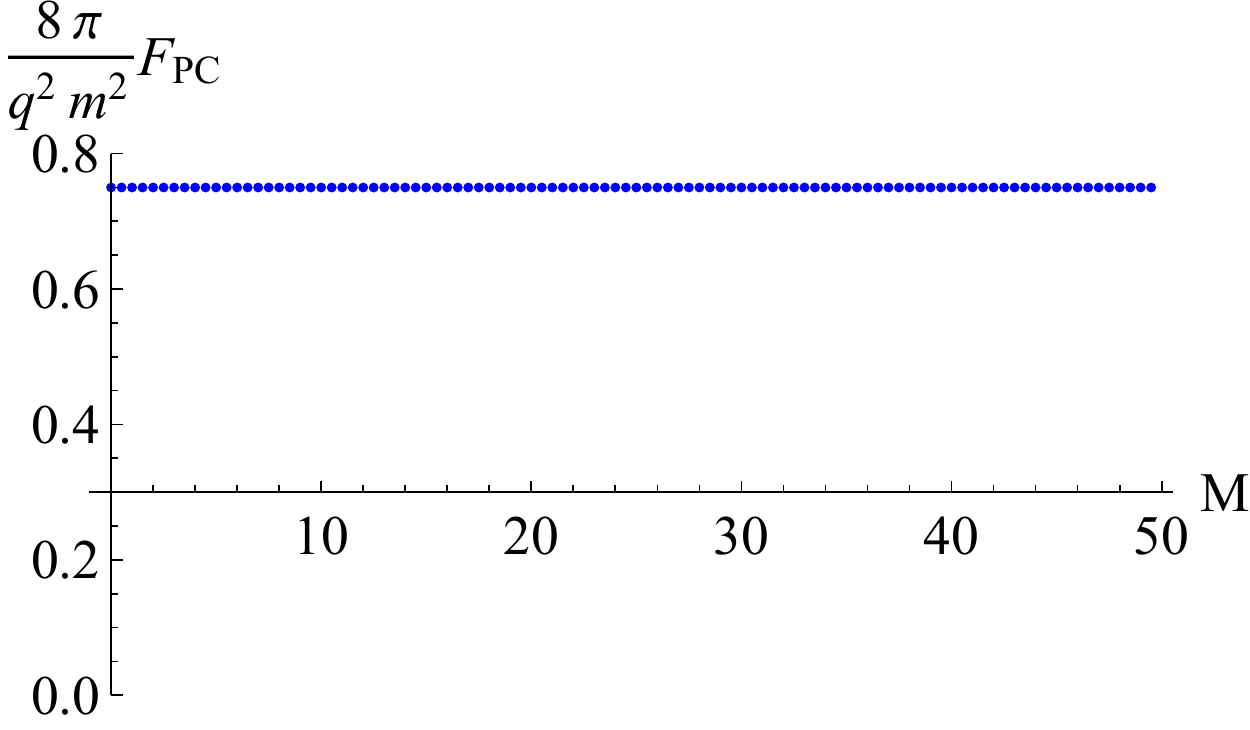} 
   \caption{Numerical calculations for the force between the Dirichlet plane and the point-like scalar charge evaluated in the limit $R=0$ and multiplied by $8\pi/(q^{2}m^{2})$ as a function of $M$.}
  \label{forcanaplaca}
\end{figure}

One could ask if the force between the Dirichlet plane and the point-like charge, multiplied by $8\pi/(q^{2}m^{2})$ and evaluated in the limit $R\to0$ would be always equal to $3/4=0.75$, independent of the value of $m$. We were not able to show that analytically, but we checked this point numerically for many other values of $M$. In figure (\ref{forcanaplaca}) we have numerical plot for the force between the Dirichlet plane and the scalar charge, multiplied by $8\pi/(q^{2}m^{2})$, and evaluated in the limit $R=0$ for several values of $M$. We can see that in all cases we have the same value $0.75$ on the vertical axis for $mR\to0$.   

\section{The case where the charge placed on the Dirichlet plate}
\label{IV}

Up to now, we have considered just the limit $R\to0$ of the energy and interaction force between the Dirichlet plane and the scalar charge. In this section we consider the interaction energy and the interaction force when the charge is placed just on the Dirichlet plane. It corresponds to take $R=0$. So, let us start by the energy.

We can also solve exactly the integral in (\ref{Energy3}) when $R=0$, the result is
\begin{eqnarray}
\label{EnergyR0}
E_{PC}\left(R=0\right)&=&\frac{q^{2}}{8\pi\sqrt{1-\frac{4M^{2}}{m^{2}}}}\int_{0}^{\infty}dp \ p \frac{\sqrt{p^{2}+m_{+}^{2}}-\sqrt{p^{2}+m_{-}^{2}}}{\sqrt{p^{2}+m_{-}^{2}}\sqrt{p^{2}+m_{+}^{2}}}\nonumber\\
&=&\frac{q^{2}\left(m_{+}-m_{-}\right)}{8\pi\sqrt{1-\frac{4M^{2}}{m^{2}}}} \ .
\end{eqnarray}

\begin{figure}[!h]
 \centering
   \includegraphics[scale=0.41]{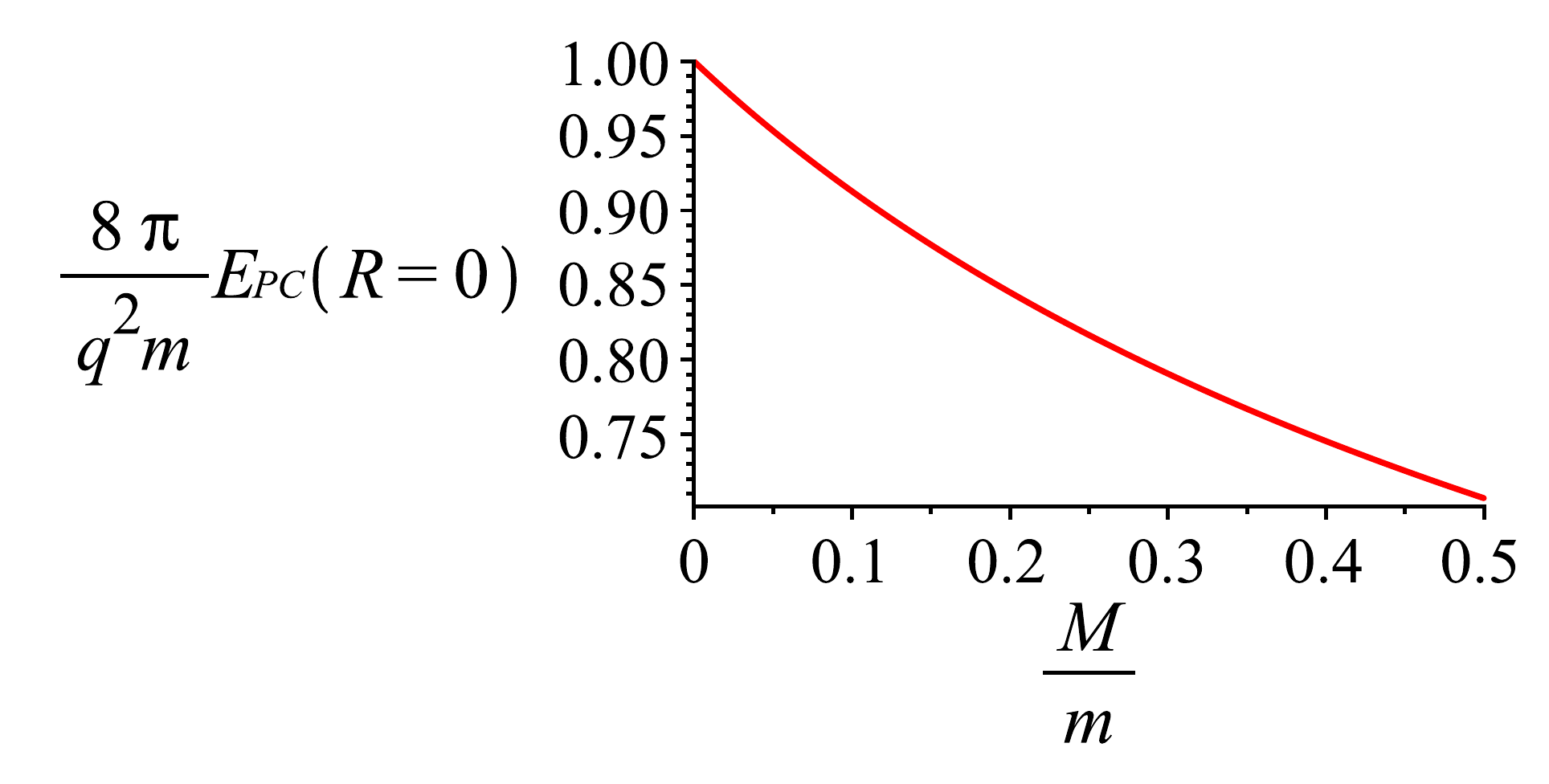} 
   \caption{Energy (\ref{EnergyR0}) multiplied by $8\pi/(q^{2}m)$ for $R=0$.}
  \label{energiaR=0}
\end{figure}

For the cases where $M=0$ and $M=m/2$ the expression (\ref{EnergyR0}) recovers the results (\ref{R0M0}) and (\ref{enrlim0}) respectively, what shows that the energy is continuous in $R=0$.  In Fig. (\ref{energiaR=0}) we have aplot for the energy (\ref{EnergyR0}) multiplied by $8\pi/(q^{2}m)$ as a function of $M/m$. The maximum value is attained when $M=0$ and the minimum, when $M/m=1/2$. 

Now let us see the behavior of the force just when the charge is placed on the plate. Taking the derivative of equation (\ref{Energy3}) with respect to $R$ and evaluating for $R=0$, we have zero as result. So interaction force between the Dirichlet plane and the scalar charge vanishes when the charge is placed on the mirror
\begin{equation}
\label{rfv1}
F_{PC}(R=0)=0\ ,
\end{equation}

According to the result (\ref{rfv1}) and the discussion of the previuos section, which suggests that $\lim_{R\to0}F_{PC}=3m^2q^2/(32\pi)$ for any value of $M$ (for the limit cases $M=0$ and $M=m/2$ it was shown analitycally), we can see that the force is finite and discontinuous on the Dirichlet plane. A discontinuity of the force on the Dirichlet plane is not properly a novel result, once the force must have its sign inverted in opposite sides of the Dirichlet mirror. In comparison with the standard Klein-Gordon field, where the force is not defined on the plane and its lateral limits diverge, for the model (\ref{Model}), the force vanishes on the Dirichlet mirror and the lateral limits are well defined (equal in modulus but with opposite signs).

Those results indicate that the higher order derivative term also tame divergences even with the presence of a Dirichlet plane.

\section{\label{V} Conclusions and final remarks}

In this paper we have investigated the scalar Lee-Wick model in the presence of a Dirichlet plane. We have obtained exactly the field propagator and we have showed that, in the limit where the Podolsky mass diverges, the propagator reduces to the one obtained from the standard Klein-Gordon field with a Dirichlet plane.

We have also studied the interaction energy, as well as the interaction force, between the Dirichlet plane and a point-like scalar charge. We have found an expression for the energy in integral form (over just a single variable), which we could solve analytically just for three specific cases. So we have made a numerical analysis. We have shown that the image method is not valid for the Lee-Wick scalar model. We have also shown that, surprisingly, when the Klein-Gordon mass vanishes, the interaction force between the Dirichlet plane and the scalar charge is the same one (with an opposite sign) obtained previously in the literature for the Lee-Wick Electrodynamics, when we have a perfectly conducting plane and an electric charge.

We have shown that the interaction energy is always finite, even when the charge is placed on the Dirichlet plate. 
Our results suggest that the interaction force between the Dirichlet plate and a point-like scalar charge multiplied $8\pi/(q^{2}m)$, and evaluated when the charge is infinitelly near to the Dirichlet plane (but not placed on it), is independent of $M$ and $m$ and is equal to $3/4=0.75$. This fact has been shown analytically for two cases, and numerically for many others. We have shown analytically that the force evaluated where the charge is placed just on the Dirichlet plane is equal to zero. So the force is always finite, but discontinuous on the mirror.

Maybe the value of 3/4 mentioned above is related to the electromagnetic mass of a point-like charge, mentioned in references \cite{LW8,Frenkel2}.

It would be interesting to consider the Lee-Wick scalar field in the presence of a semi-transparent Dirichlet plane and investigate how some typical divergences found for the scalar field would, or would not, be tamed with this kind of surface \cite{andamento}. Some preliminar results suggest, among other non trivial results, that the force is still zero but becomes continuous on the Dirichlet plane, \cite{EversonTese} .

\ 

\textbf{Acknowledgments.}  L.H.C. Borges and E.H. Rodrigues thank to CAPES (Brazilian agency) for financial support.  F.A. Barone thanks to CNPq (Brazilian agency) under the Grants 311514/2015-4 and  313978/2018-2 for financial support.

\end{document}